\documentclass[useAMS,usenatbib,usegraphicx]{mn2e}

\newcommand{\BP}{Ballesteros-Paredes}

\newcommand{\bu}{{\bmath v}}
\newcommand{\cs}{c_{\rm s}}

\newcommand{\Eg}{E_{\rm grav}}
\newcommand{\Ek}{E_{\rm kin}}

\newcommand{\kms}{{\rm ~km~s}^{-1}}
\newcommand{\LJ}{L_{\rm J}}
\newcommand{\mass}{{\cal M}}
\newcommand{\Ms}{M_{\rm s}}
\newcommand{\pcc}{{\rm ~cm}^{-3}}

\newcommand{\sfrff}{SFR$_{\rm ff}$}
\newcommand{\tff}{t_{\rm ff}}
\newcommand{\tgrav}{t_{\rm grav}}
\newcommand{\tturb}{t_{\rm turb}}
\newcommand{\vrms}{v_{\rm rms}}
\newcommand{\VS}{V\'azquez-Semadeni}

\title[The Velocity Field in Molecular Clouds]{The Nature of the
Velocity Field in Molecular Clouds. I. The Non-Magnetic Case} 
\author[\VS\ et al.] {Enrique \VS$^{1}$ \thanks{E-mail:
e.vazquez@astrosmo.unam.mx}, Ricardo F. Gonz\'alez$^{1}$ \thanks{E-mail:
rf.gonzalez@astrosmo.unam.mx}, Javier \BP$^{1}$ \thanks{E-mail:
j.ballesteros@astrosmo.unam.mx},
\newauthor  Adriana Gazol$^{1}$ \thanks{E-mail:
a.gazol@astrosmo.unam.mx}, and Jongsoo Kim$^2$\thanks{E-mail:
jskim@kasi.re.kr}\\ 
$^{1}$Centro de Radioastronom\'\i a y Astrof\'\i sica,
Universidad Nacional Aut\'onoma de M\'exico, Apdo. Postal 3-72, Morelia,
58089, M\'exico\\
$^{2}$ Korea Astronomy and Space Science Institute, 61-1,
Hwaam-dong, Yuseong-gu, Daejon 305-764, Korea}

\begin{document}



\maketitle

\label{firstpage}

\begin{abstract}

We present numerical simulations designed to test some of the hypotheses
and predictions of recent models of star formation. We consider a set of
three numerical simulations of randomly driven, isothermal,
non-magnetic, self-gravitating turbulence with different rms Mach
numbers $\Ms$ and physical sizes $L$, but with approximately the same
value of the virial parameter, $\alpha \approx 1.2$. { We obtain the
following results: a) We test the hypothesis that the collapsing centers
originate from locally Jeans-unstable (``super-Jeans''), subsonic
fragments; we find no such structures in our simulations, suggesting
that collapsing centers can arise also from regions that have supersonic
velocity dispersions but are nevertheless gravitationally unstable. b)
We find that the fraction of small-scale super-Jeans structures is
larger in the presence of self-gravity. c) There exists a trend towards
more negative values of the velocity field's mean divergence in regions
with higher densities, implying the presence of organized inflow motions
within the structures analysed. d) The density probability density
function (PDF) deviates from a lognormal in the presence of
self-gravity, developing an approximate power-law high-density tail, in
agreement with previous results. e) Turbulence alone in the large-scale
simulation ($L=9$ pc) does not produce regions with the same size and
mean density as those of the small-scale simulation ($L=1$ pc). Items
(b)-(e) suggest that self-gravity is not only involved in causing the
collapse of Jeans-unstable density fluctuations produced by the
turbulence, but also in their {\it formation}. }

We then measure the ``star formation rate per free-fall time'', \sfrff, as a
function of $\Ms$ for the three runs, and compare with the predictions
of recent semi-analytical models. We find marginal agreement to within
the uncertainties of the measurements. However, within the $L = 9$ pc
simulation, subregions with similar density and size to those of the $L
= 1$ pc simulation differ qualitatively from the latter in that they
exhibit a global convergence of the velocity field $\nabla \cdot \bu
\sim -0.6 \kms$ 
pc$^{-1}$ on average. { This suggests that the assumption that turbulence
in clouds and clumps is purely random is incomplete.}  We conclude that
a) part of the observed velocity dispersion in clumps must arise from
clump-scale inwards motions, even in driven-turbulence situations, and
b) analytical models of clump and star formation need to take into
account this dynamical connection with the external flow and the fact
that, in the presence of self-gravity, the density PDF may deviate from
a lognormal.

\end{abstract}
 

\section{Introduction} \label{sec:intro}

The role of the velocity field in the process of star formation (SF)
remains not fully understood. It is generally believed nowadays that
supersonic {turbulent motions of a given size scale} in molecular
clouds (MCs) should have a dual role in relation to SF \citep[e.g.,
][]{Sasao73, BVS99, VP99, KHM00, MO07}: {they are thought to provide
support towards scales larger than their own, while simultaneously
promoting collapse of smaller scales}. However, the origin and nature of
the observed supersonic motions remains controversial. More than three
decades ago,
\citet{GK74} suggested that the observed supersonic widths of various
molecular lines could be
representative of gravitational contraction, although this suggestion
was dismissed shortly thereafter by \citet{ZP74} through the argument
that if all MCs were collapsing and converting most of their mass into stars
in roughly one 
free-fall time, the resulting star formation rate would be at least 10
times larger than that presently observed in the Galaxy. \citet{ZE74}
then proposed that the supersonic linewidths in the clouds are produced
primarily by local, { small-scale motions, a scenario to which we refer} as ``local turbulence'',
and which has been widely accepted until recently. { Built into this
scenario is the notion that local turbulence acts as a sort of isotropic
pressure, so that it provides an important (and perhaps dominant)
contribution in the support of the clouds against
their self-gravity, in particular allowing for a state of near 
hydrostatic equilibrium \citep[e.g.,][]{MT03, HC08, FBK06}.}

However, this { scenario} faces a number of problems. First, it is
well known \citep[e.g., ][]{Frisch95} that turbulent flows possess the
largest velocity differences at the largest scales, a property {
which implies that the largest velocity differences within a turbulent
structure are expected to occur at scales comparable to the size of the
structure itself. That is, Lagrangian (i.e., moving with the flow)
clumps are expected to be continually distorted (sheared and/or
compressed) by these large-scale motions, in contradiction with the idea
of their being in near equilibrium \citep{BVS99}. Such large-scale
motions are indeed observed in molecular clouds and their
substructure \citep[e.g.,][]{Brunt02, OM02, HeBr07}}.

Second, there is the problem of how
to maintain the observed turbulence levels. Early suggestions were that
the turbulent motions consisted of hydromagnetic waves \citep[e.g.,
][]{AM75, Mousch76, SAL87}, which would be less dissipative than supersonic
hydrodynamic turbulence. However, numerical experiments demonstrated
that MHD turbulence generally decays as fast as hydrodynamic turbulence
\citep{SOG98, ML99, PN99}, except perhaps if it is unbalanced
\citep[i.e., if the energy flux along field lines in one direction differs from
that in the opposite direction;][]{CLV02}, although it is not clear
whether this applies to MCs.  

Maintenance of the turbulence in MCs by stellar energy
feedback has also been proposed, although it is not yet clear whether
this feedback can keep the clouds near equilibrium \citep[e.g.,
][]{NS80, McKee89, MM00, KMM06, NL07} or rather disrupt them \citep[e.g.,
][]{Whitworth79, Larson87, FST94, HBB01}. In both cases, the stellar
feedback has been proposed as a regulating mechanism of the star
formation efficiency (SFE). In the former, because the high stationary
turbulence levels are expected to maintain the star formation rate
(SFR) at low values \citep[e.g., ][]{KHM00, HMK01, VBK03, VKB05, NL07}, while
in the latter the
SFR can be large after a cloud forms, but the cloud is soon dispersed
after SF begins. Indeed, clusters over 5-10 Myr old are generally
observed to be devoid of molecular gas \citep[e.g., ][]{LBT89, HBB01, BH07}.

Another alternative that has been proposed recently is that MCs obtain their
turbulence from the compressions that form them in the first place
through various instabilities
\citep{KI02, VBK03, AH05, Heitsch_etal05, Heitsch_etal06, VS_etal06},
although in this case the turbulence should begin to decay after the
compression that formed the cloud subsides. \citet{VS_etal07} showed
simulations in which, when this happens, the cloud begins to contract
gravitationally, exhibiting a virial-like energy balance $|E_{\rm grav}|
\sim 2 E_{\rm kin}$ while doing so, even though the cloud is never in virial
equilibrium. { This can in fact be understood because, in the case of
gravitational collapse, the velocity dispersion is at most within a
factor of $\sqrt{2}$ from the value needed for equipartition.  Thus,
both a system in virial equilibrium and one in free-fall are, to order
of magnitude, in energy equipartition, and are, thus, observationally
indistinguishable on the basis of velocity dispersion alone \citep{BP06}.}

If the simulations by \citet{VS_etal07} are representative of actual 
MCs, the nonthermal motions in the clouds 
could transit from being initially due to actual random turbulence, to
later being due to gravitational contraction. In this case,
the pseudo-virial energy balance would be a
manifestation of the gravitational contraction rather than of virial
equilibrium. We refer to this
mode of flow, in which the velocity field at all scales contains a
significant inflow component, as {\it large-scale inflow (LSI)}. Such a
flow can 
be due to either generic dynamic compressions in the ISM (e.g., expanding HII
regions or supernova remnants, or the transonic turbulence in the warm
ISM), or to various large-scale instabilities, such as gravitational
or magneto-Jeans \citep[e.g., ][ see also the review by
 Hennebelle, Mac Low \& \VS\ 2008]{LMK05, KO06}.

The scenario of LSI driven by self-gravity is actually
frequently encountered. Since it is standard in simulations of
star formation in 
clouds with decaying turbulence \citep[e.g., ][]{BBB03, BB06}, it is
often associated with a regime of decaying turbulence, although in
principle there is no reason why it should be only applicable
in this case. \cite{FBK06}
have attempted to describe a gravitationally-driven mass cascade { that
involves the formation of smaller-scale structures by gravitational
contraction of larger-scale ones, followed by virialization at the small
scales. The LSI scenario} is also consistent with a number of recent studies 
suggesting that indeed some MCs \citep{HaBu07} and clumps \citep{PHA07}
may be undergoing global gravitational contraction. 
It should be noted, however, that the simulations supporting this
scenario have generally 
been non-magnetic, possibly biasing the results. \citet{Elm07}
has recently suggested that clouds may collapse in regions where they
are magnetically supercritical, while their subcritical fragments may
remain supported for times significantly longer than their free-fall
time.

One fundamental distinction between the hypotheses of local turbulence
and of LSI is that, in the former, the kinetic energy of the turbulent
motions internal to a clump is assumed to act fully as support against
gravity, while in the latter, part of the kinetic energy of these motions
may be globally compressive, either promoting collapse or being a
consequence of it \citep{HF82, BVS99, BP06, Dib_etal07}. Recent models
of the SFR \citep[e.g., ][]{Elm02, KM05} or of the turbulent
clump mass function as the origin of the stellar initial mass function
\citep{PN02, HC08} { have been formulated under the assumption of local
turbulence. It is thus important to determine whether
this assumption is verified in numerical simulations of self-gravitating
turbulence. The realization of the LSI scenario would thus imply a
reduced amount of support against self-gravity.}

{ Moreover, the models by \citet[][hereafter PN02]{PN02}, \citet{Elm02} and
\citet[][hereafter KM05]{KM05} rely on the idea advanced by 
\citet{Padoan95}, \citet{Elm02} and \citet{VBK03}} that the mass that
proceeds to collapse is that which is deposited by the turbulence in
subsonic, yet Jeans-unstable (``super-Jeans'') fragments. The latter
authors provided indirect evidence that this could be so by showing the
existence of a correlation between the SFE and the so called ``sonic
scale'' of the turbulence, the scale below which the turbulent motions
are subsonic on average. However, it is not necessary that {\it only}
the mass in these subsonic, super-Jeans structures proceeds to
collapse. Material in supersonic, yet effectively gravitationally
unstable structures can also participate in the collapse.

The goal of the present paper is to contribute towards the understanding
of the role of the velocity field's topology on the control of the star
formation process. To this end, we use numerical simulations (described
in\S \ref{sec:model}) of randomly driven turbulence in isothermal,
self-gravitating flows aimed at investigating whether the hypotheses and
predictions of analytical models are verified in the non-magnetic
case. The simulations have different sizes, mean densities and velocity
dispersions, but scaled so that all have approximately the same value of
the virial parameter $\alpha \equiv 2 \Ek /|\Eg|$, thus satisfying the
hypotheses of the KM05 
model. In these simulations, we first search for the fraction of
super-Jeans, subsonic subregions in a cloud, to see whether they can be
deemed responsible for the mass that ends up collapsing (\S
\ref{sec:subs_superJ}). Second, we investigate whether the internal
velocity dispersion in dense subregions of a cloud can be considered as
random as that at larger scales (\S \ref{sec:div_v}), or instead
exhibits increasing amounts of the compressive component as the density
of the structures increases and their size decreases. In \S
\ref{sec:SFE}, we ask whether the suite of simulations agrees with the
prediction of the model by KM05 for the dependence of the \sfrff\ on the
turbulent Mach number. Finally, in \S \ref{sec:discussion} we summarize
and discuss the implications of our results, in particular comparing
with previous work.

\section{The models} \label{sec:model}

We have performed three simulations of non-magnetic, self-gravitating,
isothermal turbulence at a resolution of $512^3$ zones, using a total
variation diminishing (TVD) scheme \citep{Kim_etal99} with { periodic
boundaries and} random Fourier driving with a spectrum $P(k) = k^6
\exp(-8 k/k_{\rm pk})$. { Here, $k$ is the wavenumber and $k_{\rm pk}
= 2 (2 \pi/L)$ is the energy-injection wavenumber, with $L$ being the
computational box size. The energy is thus injected mostly at scales of
order half the box size.} The driving is purely rotational (or
``solenoidal''), thus having no compressive component. A prescribed rate
of energy injection is applied in order to approximately maintain the
rms Mach number $\Ms \equiv \sigma/\cs$ near a ``nominal'' value that
characterizes the run. Here, $\sigma$ is the three-dimensional velocity
dispersion and $\cs$ is the sound speed, taken equal to $0.2 \kms$ in
all runs (corresponding to $T = 11.4$ K).  The actual value of $\Ms$
fluctuates and is slightly different from the nominal value, because the
numerical scheme is designed to maintain a constant energy injection
rate, not a constant rms Mach number. 

The other parameter that
characterizes a simulation is the ``Jeans number'' $J \equiv L/\LJ$,
where $\LJ
\equiv (\pi \cs^2/G\rho)^{1/2}$ is the Jeans length, with $\rho \equiv
\mu n_0$ being the mean density of the simulation, $n_0$ the mean number
density, $\mu = 2.36~m_{\rm H}$ the mean particle mass, and $m_{\rm H}$
the mass of the Hydrogen atom. All three simulations are
evolved for 3.2 turbulent crossing times before turning on 
self-gravity, in order for the turbulence to reach a fully developed
state, and thus avoid applying the self-gravity directly on the imprints
of the random driving.
The three simulations differ in physical size $L$, mean density $n_0$
and nominal rms Mach number $\Ms$, but in each case their values are
chosen as to give the same value of the ratio $\Ms/J$. The nominal
values of the pairs $(\Ms,J)$ for the three runs are $(8,2)$, $(16,4)$,
and $(24,6)$, respectively, corresponding to a nominal value of $\Ms/J =
4$. { The runs are named mnemonically by means of their values of $\Ms$
and $J$.}

Note that $(\Ms/J)^2$ is proportional to the virial parameter $\alpha$,
as can be seen by approximating $\Ek \approx \mass \sigma^2/2$, where
$\mass$ is the total mass in the simulation, and $|\Eg| \approx
G\mass^2/L$. For a spherical cloud of mass $\mass =4 \pi \rho L^3/3$, we
thus have
\begin{equation}
\alpha \equiv \frac{2 \Ek}{|\Eg|} \approx \frac{\mass \sigma^2}{G \mass^2/L} = 
\frac{3}{4\pi^2} \frac{\Ms^2}{J^2}.
\label{eq:alpha_M/J}
\end{equation}
Our simulations thus all have a nominal value of $\alpha
\approx 1.22$. The actual average values of $\Ms$ and of $\alpha$,
together with other 
parameters of the runs, are indicated in Table
\ref{tab:run_parameters}, { in which the column labeled $\tgrav$
gives the time at which self-gravity is turned on, and the column
labeled ``$\Ms$ (real)'' gives the actual measured average value of the
rms Mach number over the duration of the run.}
\begin{table*}
\begin{center}
\vskip 0.5cm
\centerline{\sc Run parameters}
\vskip 0.2cm
\begin{tabular}{cccccccccccccccccc}
\hline
\hline
\noalign{\smallskip}
Name &  $L$  & $n_0$ & {\cal M} & 
$\LJ$ &  J &  $\Ms$ (real)  & $\alpha$ (real) & $v_{\rm rms}$  &
$\tff$ &  $\tgrav$ & $\tturb$ (nominal) & grid cell size\\
& [pc] & [cm$^{-3}$] & [ $M_{\odot}$] &  [pc] & & & & [$\kms$] & [Myr]  &
[Myr] & [Myr] & [pc] \\
\hline
\noalign{\smallskip}
Ms8J2   & 1 & 2000   & 115.8& 0.5 & 2 & 8.6 & 1.4 & 1.7 & 2.5 & 2 & 0.625 & 0.00195\\
Ms16J4  & 4 & 500    & 1853 & 1 & 4 & 15.7 & 1.2 & 3.1 & 5 & 4 & 1.25 & 0.00781\\
Ms24J6  & 9 & 222.22 & 9382 & 1.5 & 6 & 23.0 & 1.1 & 4.6 & 7.5 & 6 & 1.875 & 0.0175\\

\noalign{\smallskip}
\hline
\label{tab:run_parameters}
\end{tabular}
\end{center}
\end{table*}

It is also worth noting that fixing the ratio $\Ms/J$ only fixes the
ratio $\sigma/(\rho^{1/2} L)$ (at a given $\cs$), and so we still have
freedom to choose the values of the individual physical parameters. We
do this by assuming that the size $L$ and mean density $n_0$ of our
simulations satisfy one of Larson's (1981) relations, i.e., $n_0
\propto L^{-1}$. Since all three simulations have the same nominal value
of $\alpha$, then the above assumption also implies that our suite of
simulations also satisfies the other Larson relation, $\sigma \propto
L^{1/2}$. The set of physical values for the simulations are also
reported in Table \ref{tab:run_parameters}.

\section{Fraction of subsonic, super-Jeans structures} \label{sec:subs_superJ}

\subsection{Procedure} \label{sec:subs_superJ_procedure}

In this section, we measure, as a function of region size, the fraction
of regions in the numerical simulations that is both subsonic and
super-Jeans, in order to test the hypothesis that these are indeed the
structures that collapse gravitationally to form stars. For generality,
we consider two types of regions: a) the set of all cubic sub-boxes of a
simulation of a given size and b) dense clumps defined by a density
threshold criterion. The first set contains both overdense and
underdense regions, { and is intended to provide unbiased statistics
of the velocity field in all subregions of a given size within the
numerical box, independently of the local density}. The second set
contains only overdense clumps.

To isolate the effect of self-gravity, we perform the
procedure at two different times in each simulation. First, we consider
the last data dump before gravity is turned on, at which
the density distribution must be a consequence of the turbulent flow
alone. Second, we consider a data dump at around two free-fall
times $\tff$ after having turned gravity on, at which significant gravitational
collapse has occurred, and the density structure should be the result of the
combined effects of turbulence and self-gravity. 

For the analysis using sub-boxes of the simulation, we subdivide the
latter in cubic sub-boxes of sizes 2, 4, 8, 16, 32, 64, and 128 grid
zones per dimension. Since the simulations are performed at a resolution
of 512 grid cells per dimension, and runs Ms8J2, Ms16J4 and Ms24J6
respectively have sizes $L = 1$, 4 and 9 pc, the grid cell size differs
for each run, and is also indicated in Table \ref{tab:run_parameters}.
{ It should be noted that, since the sub-boxes are located at fixed
positions within the simulation box, their locations bear no
special relation with those of actual clumps. A clump may be located at
the interface between two sub-boxes, and it in general moves from one
sub-box to another as time elapses. However, we make no attempt to
repeat the procedure shifting the positions of the sub-boxes to find a
possible ``best'' match between the sub-boxes and the actual clumps. We
consider that the number of sub-boxes is large enough that no particular
choice of the origins of the boxes is better than any other.}

For the analysis with clumps, we define a clump as a connected set of
grid cells with densities above a certain threshold. To create an
ensemble of clumps, we consider a series of thresholds, of $n =$ 32, 64,
128 and 256 times the mean density of the simulation. { We then
approximate their size by $L = (3V/4 \pi)^{1/3}$, where $V$ is their
volume. As discussed by
\citet{VKSB05}, this is a rather robust estimator of the clumps'
size. Finally, for the plots, the clumps are classified by their sizes
(independently of the threshold from which they originated) and added to
logarithmic size bins defined by successive powers of 2 from 2 to
64 grid cells.}

For each region {(sub-box or clump)}, we compute its mean density,
its internal three-dimensional velocity dispersion $\sigma^2$
(substracting the region's mean velocity), and whether it is Jeans
stable or unstable. In the case of sub-boxes, since their size is
predetermined { and the temperature is constant in space and time},
we simply compute the critical density for { Jeans} instability {
(i.e., with respect to thermal support only)\footnote{{ We do not
consider the non-thermal contribution to the ``support'', since we are
only interested in the fraction of simultaneously subsonic and
super-Jeans structures. Moreover, as discussed in \S \ref{sec:div_v}, it is not
clear that all of the non-thermal kinetic energy can be considered to
provide support.}}} at the specified
size. { Since the flow is isothermal, this ``Jeans density'' is
unique at a given scale, and thus we can 
compare it with the sub-box's mean density.}  For the clumps, { we simply compare their
estimated size to the Jeans length derived from the clump's mean
density.}
%
%
Finally, we count the number of regions that have a size larger than 
their associated Jeans length, the number of regions that
have a subsonic velocity dispersion, and the number of regions that
satisfy both conditions simultaneously.

\begin{figure*}
\includegraphics[width=1.\hsize]{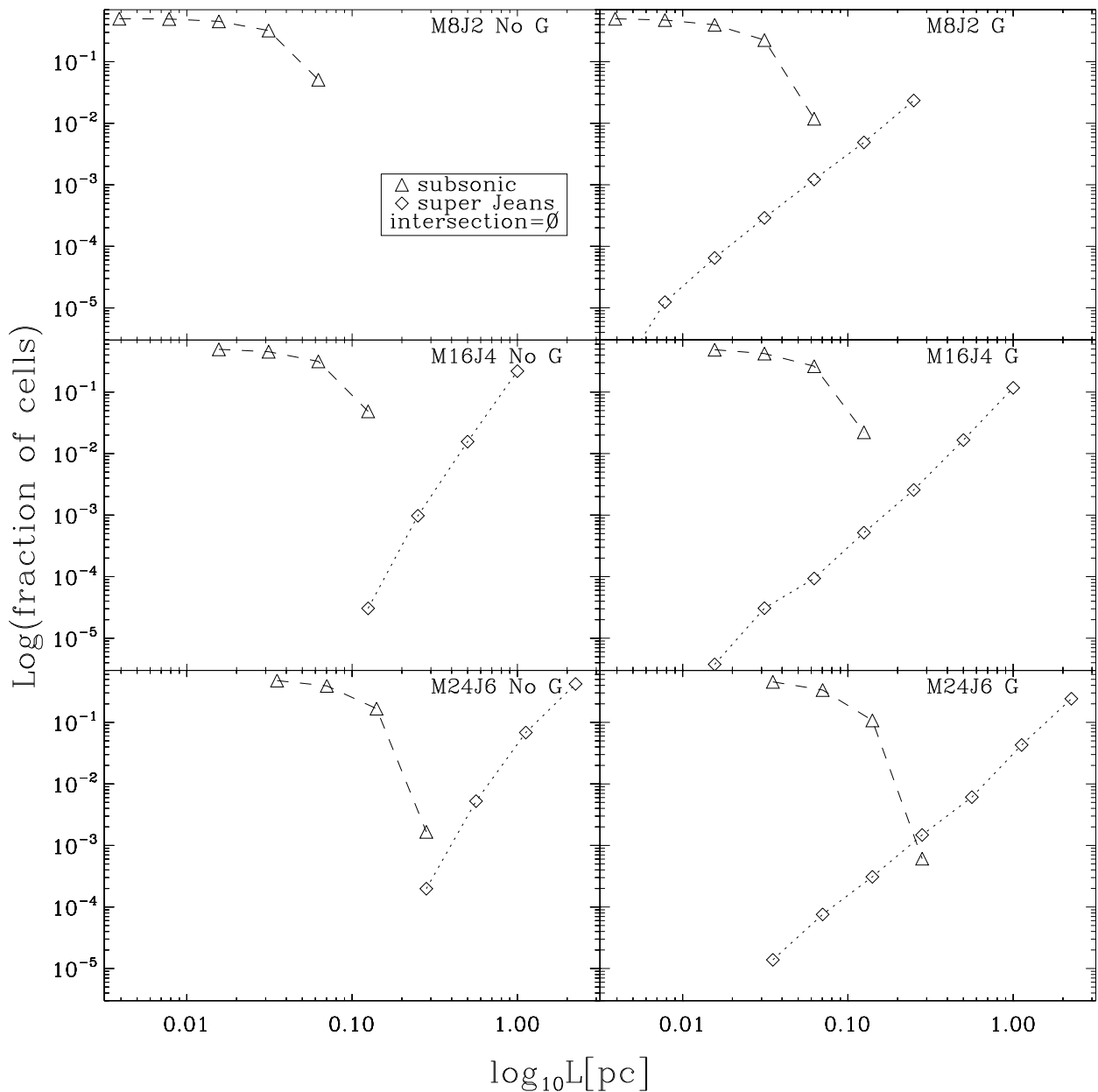}
\caption{Fraction of subsonic ({\it triangles, dotted lines}) and of
super-Jeans ({\it diamonds, solid lines}) sub-boxes in runs Ms8J2 ({\it top
row}), Ms16J4 ({\it middle row}), and Ms24J6  ({\it bottom
row}), as a
function of sub-box size. The entire numerical box is subdivided in sub-boxes
of the indicated size. The {\it left} panels show the fractions shortly
before the time when self-gravity is turned on ($\tgrav$). The
{\it right} panels show the fractions at approximately two free-fall
times after $\tgrav$. The fraction of sub-boxes that are
both subsonic and super-Jeans is zero at all sub-box sizes, and thus
cannot be shown in this figure.}
\label{fig:sprJ_subson_cells}
\end{figure*}

\begin{figure*}
\includegraphics[width=1.\hsize]{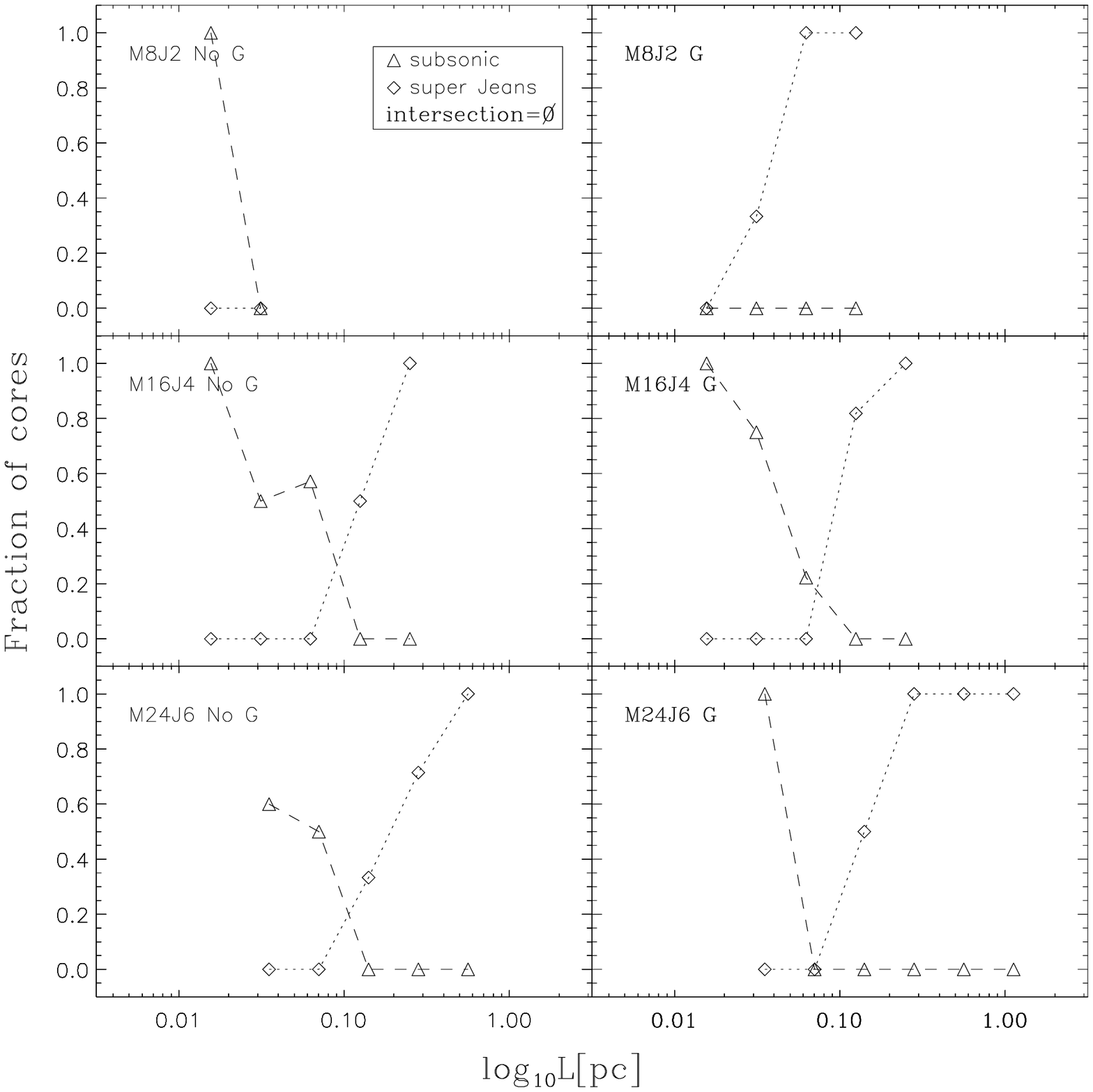}
\caption{Same as Figure \ref{fig:sprJ_subson_cells} but for clumps rather
than sub-boxes. The clumps are defined as connected structures with densities
above a density threshold. The ensemble of clumps was created by
considering thresholds 32, 64, 128 and 256 times the mean density
$n_0$. The fraction of clumps that are both subsonic and 
super-Jeans is again zero at all clump sizes considered (not
shown). Scales at which there are no clumps have no points drawn.}
\label{fig:sprJ_subson_cores}
\end{figure*}

Note also that, in the case of clumps, we have not corrected for
the possibility of them being located at the boundaries of the periodic
numerical box. So, if a clump crosses one or more boundaries, it is
artificially split into two or more fragments by our clump-defining
algorithm. However, we do not expect this minor omission to introduce
any serious biases since, { as we discuss in \S
\ref{sec:subs_superJ_results}}, the fraction of super-Jeans 
clumps is found to already be unity at the largest sizes (clumps of even
larger size would be even more super-Jeans), and the fraction of
subsonic clumps is already zero at scales significantly smaller than the
largest clump sizes found (clumps of larger size would be expected to be
even more supersonic). Besides, the sizes of the clumps are never larger
than 1/10th of the numerical box, and so the probability of them
crossing the boundary is relatively low.

\subsection{Results} \label{sec:subs_superJ_results}

The fractions of
subsonic and super-Jeans structures { for the three runs} are shown
for sub-boxes in Fig. \ref{fig:sprJ_subson_cells} and for clumps in Fig.
\ref{fig:sprJ_subson_cores}. In both figures, the { {\it top row}
shows results for run Ms8J2, the {\it middle row} shows run Ms16J4, and
the {\it bottom row} shows run Ms24J6;} the {\it left} panels show
the result before { the time at which self-gravity is turned on
($\tgrav$; cf. Table \ref{tab:run_parameters}), and the {\it right}
panels show it at approximately two free-fall times after $\tgrav$.}
Note that, for the sub-boxes, the fractions can be very small, and
are thus shown in logarithmic scale.

The most notable result of this analysis is
that {\it the set of 
simultaneaously subsonic and super-Jeans structures (either sub-boxes or
clumps) is empty at all the scale sizes we sampled.} Thus, this fraction
is {\it not} plotted in Figures \ref{fig:sprJ_subson_cells} and
\ref{fig:sprJ_subson_cores}. We discuss the implications and
limitations of this result in \S \ref{sec:discussion}.

Some other features are worth noting. 
The fraction of subsonic ({\it triangles, dotted lines}) sub-boxes or
clumps as a function of size shows no clear trend with the inclusion of
self-gravity.  However, the fraction of super-Jeans ({\it diamonds,
solid lines}) structures at small (for sub-boxes) or intermediate (for
clumps) scales tends to increase in the presence of self-gravity. In
fact, for the small-scale run Ms8J2 there are no super-Jeans sub-boxes
in the absence of self-gravity, but significant amounts appear after it
has been turned on.  This means that {\it the presence of self-gravity
changes the distribution of sub-box masses in comparison to that
produced by turbulence alone, increasing the fraction of regions that
can proceed to gravitational collapse.} That is, the effect of
self-gravity can begin {\it prior} to the actual ``capture'' of a region
to proceed to collapse.\footnote{{ We say that a certain density
enhancement is ``captured'' by gravity when it was originally produced
by a turbulent compression, but at some time while its mass and/or
density are increasing, it suddenly becomes gravitationally unstable and
begins to undergo gravitational collapse.}} In other words, a
density enhancement is in general expected to reach higher peak and mean
densities in the presence of self-gravity than when it is { produced}
only by turbulence. We refer to this as gravity aiding in the {\it
production} of the density fluctuation, regardless of whether it 
{ eventually collapses} or not.

\begin{figure*}
\includegraphics[width=1.\hsize]{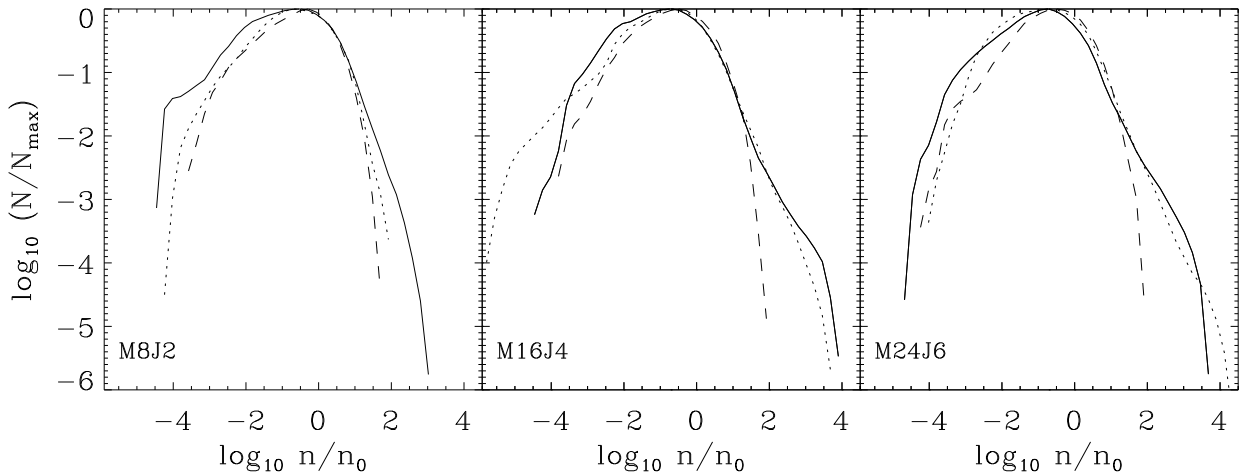}
\caption{Probability density function (PDF) of the density field in the
three runs at times $t=0~\tff$ ({\it dashed line}), $1~\tff$ ({\it
dotted line}), and $3~\tff$ ({\it solid line}) after gravity was turned 
on, where $\tff$ is the global free-fall time of each simulation
(cf. Table \ref{tab:run_parameters}). The vertical axis is normalized to
the number of grid cells at the maximum, $N_{\rm max}$.}
\label{fig:rho_PDFs}
\end{figure*}

This conclusion is also supported by the probability density function
(PDF) for the density field of { the three runs. In
Fig. \ref{fig:rho_PDFs} we show the density PDFs of the three runs at 0,
1 and 2 times the global free-fall time, $\tff$, of each simulation
(cf. Table \ref{tab:run_parameters}). The PDF at $0~\tff$ is
representative of the effects of turbulence alone, while those at 1 and
2 $\tff$ show the effect of turbulence and gravity combined. In all
runs, the PDFs in the presence of self-gravity show a prominent
high-density tail, implying that the relative frequency of high density
regions is higher in this case, compared to the effect of turbulence
alone. Similar results have been found by other workers
\citep{Klessen00, DB05}}. Thus, the {\it production} of super-Jeans
structures is itself aided by the inclusion of self-gravity.

{ One final point to note is that, since higher density thresholds in
general imply that the resulting clumps have smaller sizes and higher
densities, the small clumps are generally also denser. Thus,
Fig. \ref{fig:sprJ_subson_cores} implies that there indeed exists a
population of dense, subsonic clumps in the simulations, which, however,
are not superJeans.}


\section{Compressive component of the velocity field in dense
structures} \label{sec:div_v}

We now investigate the nature of the velocity field in subregions of our
turbulent supersonic flows, in particular aiming at whether our
small-scale simulation Ms8J2 is statistically representative of regions
of similar size within the large-scale run Ms24J6. For this purpose, we
{ again} subdivide the large-scale simulation's (Ms24J6) domain into
sub-boxes, { but this time} having the same size as the small-scale run
Ms8J2.
Note, however, that run Ms8J2 has a size of 1/9th that of run
Ms24J6. Since the latter has a resolution of $512$ grid cells per
dimension, then the sub-boxes should span $512/9 = 56.89$ cells per
dimension, which we round to 57 cells. Due to this rounding, we cannot
fit 9 full, non-intersecting sub-boxes along each direction within the
whole box, and thus we only consider the first 8 sub-boxes from the
origin in each direction, for a total of $8^3=512$ sub-boxes, { each
with a linear size of 57 grid cells.}

For each sub-box, we measure its mean density and the mean
divergence of its velocity field. To compute the divergence, we
Fourier-transform the three velocity components for the entire
simulation box, and compute the divergence in Fourier space as
\begin{equation}
{\cal F}(\nabla \cdot v) = -i {\bmath k} \cdot {\bmath v_k},
\label{eq:Four_div}
\end{equation}
where $i=\sqrt{-1}$, $\cal F()$ denotes the Fourier transform of its
argument, ${\bmath k}$ is the wavevector, and ${\bmath v_k}$ is its
associated Fourier velocity amplitude. We then transform back to
physical space to obtain the divergence { field for the whole box},
and finally we take the average of this field in each sub-box.

\begin{figure*}
\includegraphics[width=0.45\hsize] {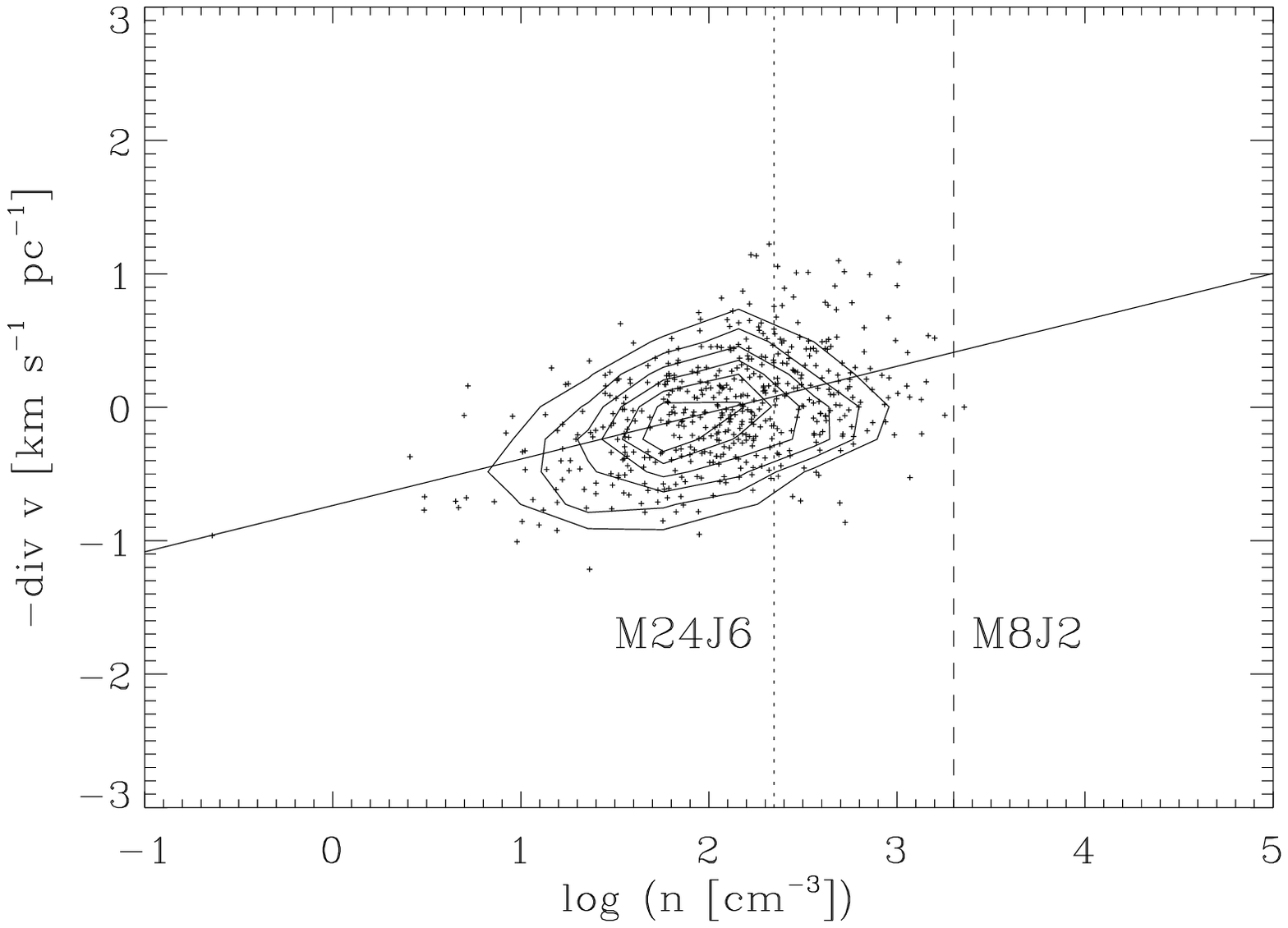} 
\hspace{0.5cm} \includegraphics[width=0.45\hsize] {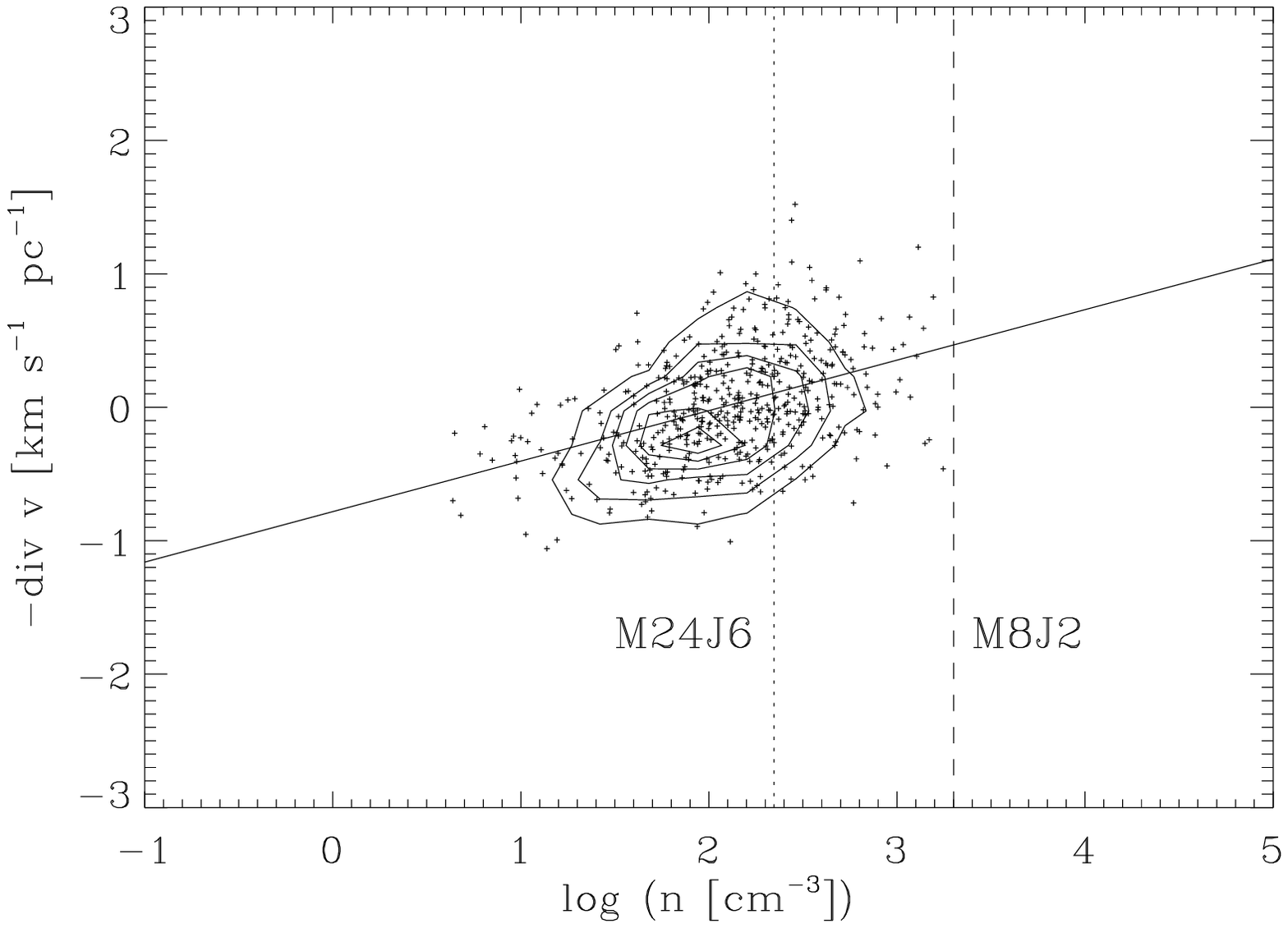} 
\caption{Negative mean divergence of the velocity field {as a function of
mean density for sub-boxes of the large-scale simulation Ms24J6 that
have the same size as run Ms8J2, at two times without self-gravity:
$t=4$ Myr, corresponding to $2.1~\tturb$ into the evolution ({\it left
panel}), and $t=6$ Myr ($3.2~\tturb$; {\it right panel}), where $\tturb$
is the turbulent crossing time of the simulation [run Ms24J6; see Table
\ref{tab:run_parameters}]). The straight {\it solid} lines show
least-squares fits through the data points, and have slopes $0.35$ ({\it
left panel}) and 0.38 ({\it right panel}). The contours show the
two-dimensional histogram of the distribution of points in the plot, at
levels 0.143, 0.286, 0.429, 0.571, 0.714 and 0.857 of the maximum. The
vertical lines show the mean density of simulations Ms24J6 ({\it dotted
line}) and Ms8J2 ({\it dashed line}). It is seen that, in the absence of
gravity, the formation of structures of the same size and mean density
as those of run Ms8J2 is extremely unlikely within run Ms24J6.}}
\label{fig:div_vs_ro_NSG}
\end{figure*}

\begin{figure*}
\includegraphics[width=0.45\hsize] {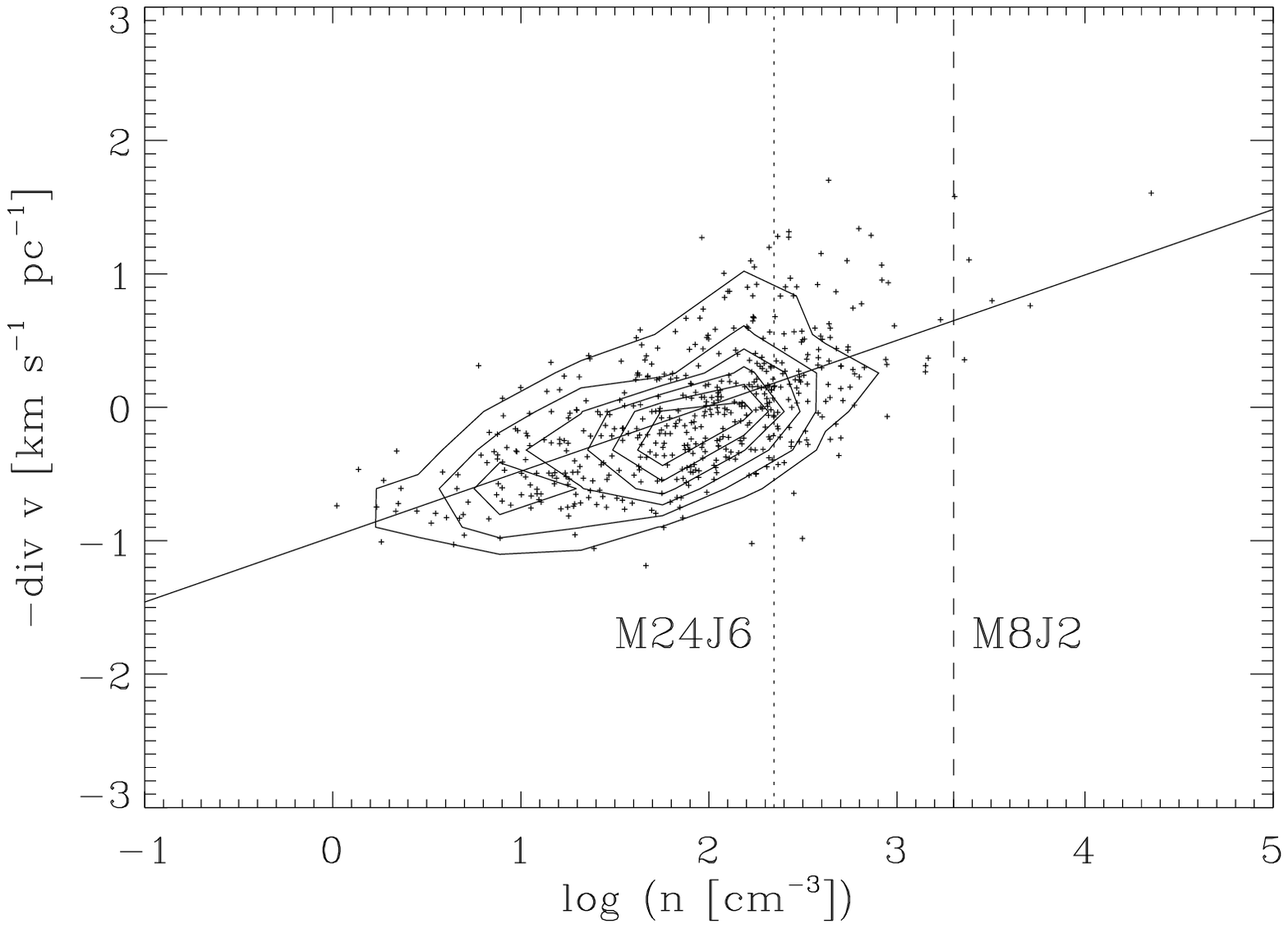} 
\hspace{0.5cm} \includegraphics[width=0.45\hsize] {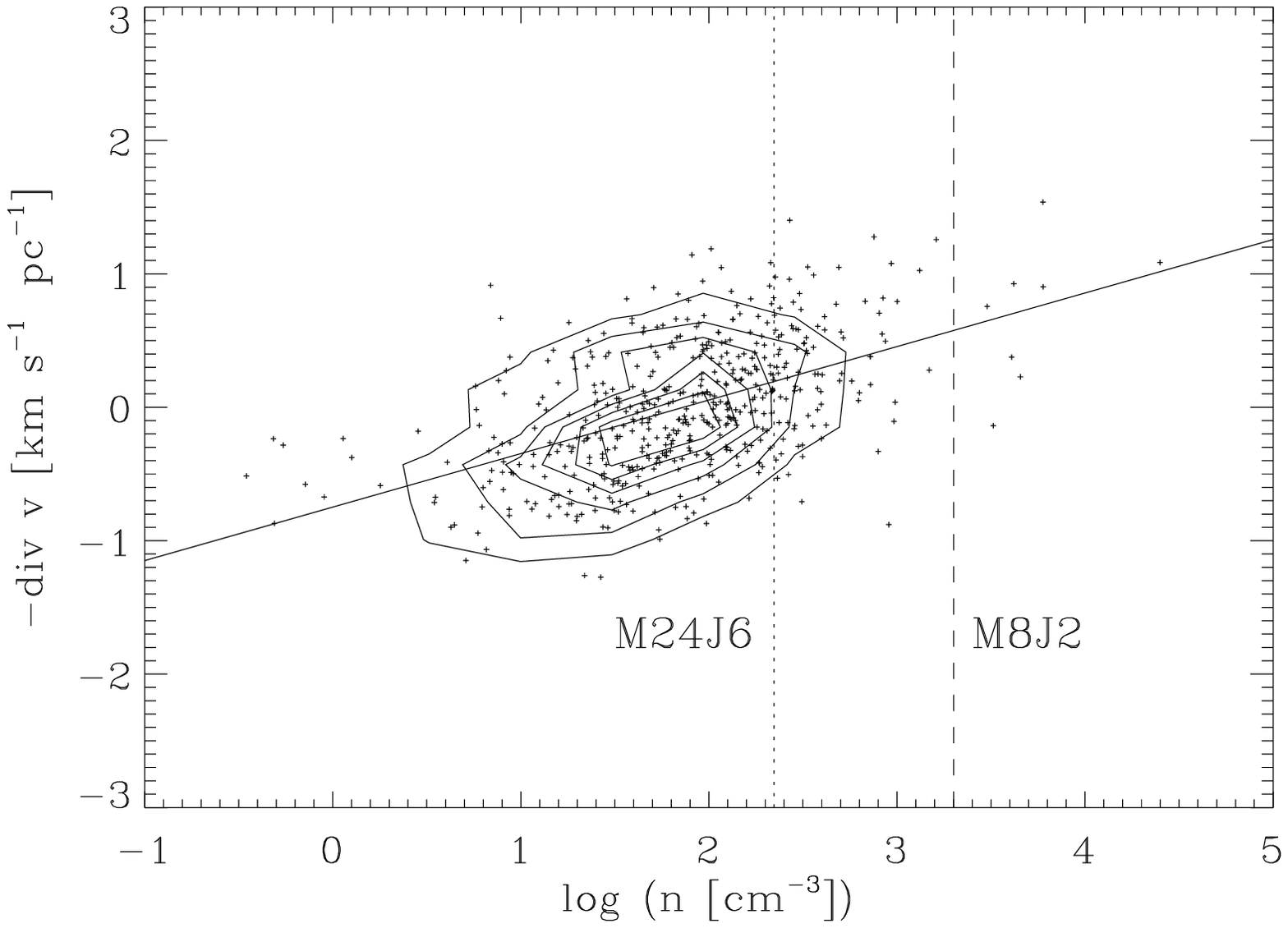} 
\caption{{ Same as Fig.\ \ref{fig:div_vs_ro_NSG} but at two times
with self-gravity: $t = 14$ Myr
({\it left panel}) and $t = 20$ Myr ({\it right panel}) (respectively
$1~\tff$ and $1.93~\tff$ after having
turned on self-gravity, with $\tff$ being the global free-fall time of
the simulation [run Ms24J6; see Table \ref{tab:run_parameters}]). The slopes
of the fits through the data points have slopes $0.50$ ({\it left
panel}) and 0.40 ({\it right panel}). We note that in the presence of
self-gravity, the likelihood of forming structures with the same size as
run Ms8J2 and mean density equal or larger than the mean density of the
latter is significantly increased with respect to the case without
self-gravity (Fig. \ref{fig:div_vs_ro_NSG}).}} 
\label{fig:div_vs_ro_SG}
\end{figure*}

Figure \ref{fig:div_vs_ro_NSG} shows the result of this exercise, giving
the mean divergence of each sub-box of run Ms24J6 as a function of its
mean density { at two times before self-gravity is turned on: at $t =
4$ Myr ($=2.1~\tturb$, where $\tturb \equiv L/\vrms
\approx 1.875$ Myr is the ``nominal'' turbulent crossing time; cf. Table
\ref{tab:run_parameters}) on the {\it left} panel, 
and at $t=6$ Myr $= 3.2~\tturb$ on the {\it right} panel. The
contours show the two-dimensional histogram of the points, and are drawn
at levels 0.143, 0.286, 0.428, 0.571, 0.714 and 0.857 of the maximum. 
Figure \ref{fig:div_vs_ro_SG} shows the corresponding plot at times in which
self-gravity has been on for $1~\tff$ ({\it left panel}) and $1.93~\tff$
({\it right panel}), where $\tff \equiv \LJ/\cs$ is the free-fall time.}
The vertical lines in both figures show the mean densities of runs
Ms24J6 ({\it dotted line}) and Ms8J2 ({\it dashed line}). 

{ In all cases, the contours are seen to have an
elongated shape, indicating that, although with abundant scatter, a
positive correlation exists between the mean value of $-\nabla \cdot
\bu$ (i.e., the velocity {\it convergence}) and the mean density of the
sub-boxes.  The contours are more elongated in the cases with
self-gravity, for which we find an average fit of}
\begin{eqnarray}
\left[\frac{\nabla \cdot \bu}{\kms {\rm pc}^{-1}}\right] &\approx& - (0.45
\pm 0.05) ~\log_{10} \left[\frac{n}{222 \pcc} \right] - \nonumber \\
&& 0.13 \pm 0.03,
\label{eq:div_vs_ro}
\end{eqnarray}
where the { uncertainties in the slope and the intercept { span} the
range of values we have encountered in the two cases sampled.}
The fit implies that, at the { size and} mean density of run Ms8J2,
which is 9 times { smaller and denser than}
run Ms24J6, a mean divergence of $\sim -0.6 \kms {\rm pc}^{-1}$
should be expected. Instead, run Ms8J2, which is driven by random
forcing at its own scale in order to mimic the assumption of local
turbulence, has zero mean divergence by construction. The trend of
$\nabla \cdot \bu$ with mean density { observed in the sub-boxes of the
large-scale simulation Ms24J6} suggests that {\it the LSI scenario} 
(cf. \S \ref{sec:intro}) {\it is verified in the
density enhancements even in driven turbulence regimes}.

It is also noteworthy that the probability of producing sub-boxes
of the same density and size as run Ms8J2 in run Ms24J6 is extremely low
(1 case in 1024, since each plot contains 512 points) in the absence of
self-gravity (Fig. \ref{fig:div_vs_ro_NSG}). This probability increases
substantially in the presence of self-gravity (13 cases in 1024). Thus,
we conclude that {\it { the formation of regions with the same size
and density as those of run Ms8J2 within run Ms24J6 
seems to require the presence of self-gravity.}}  This result further
reinforces the conclusion from \S \ref{sec:subs_superJ} that
self-gravity intervenes in the {\it formation} of super-Jeans
structures, and not only in their collapse.

\begin{figure*}
\includegraphics[width=0.45\hsize] {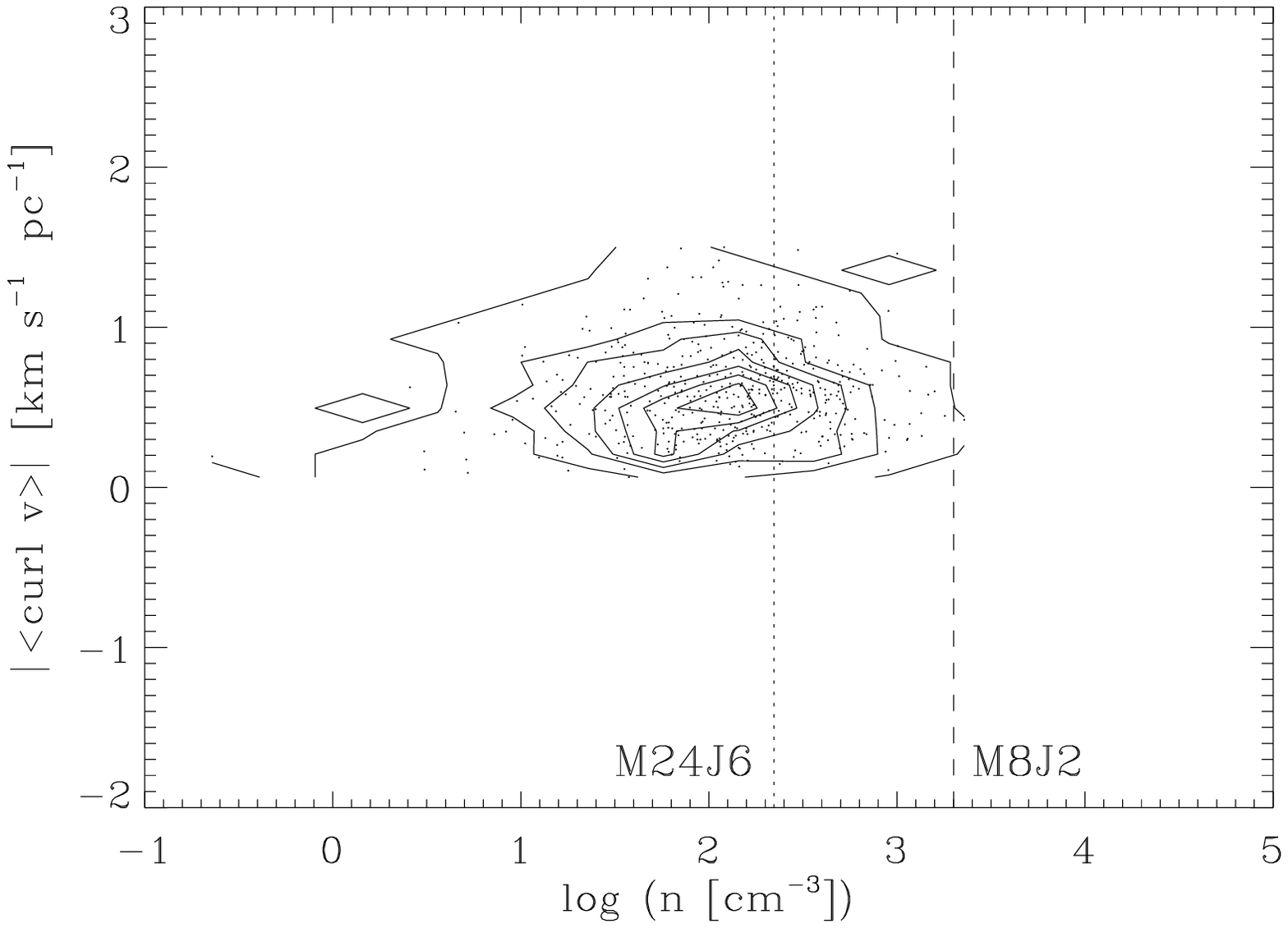} 
\hspace{0.5cm} \includegraphics[width=0.45\hsize] {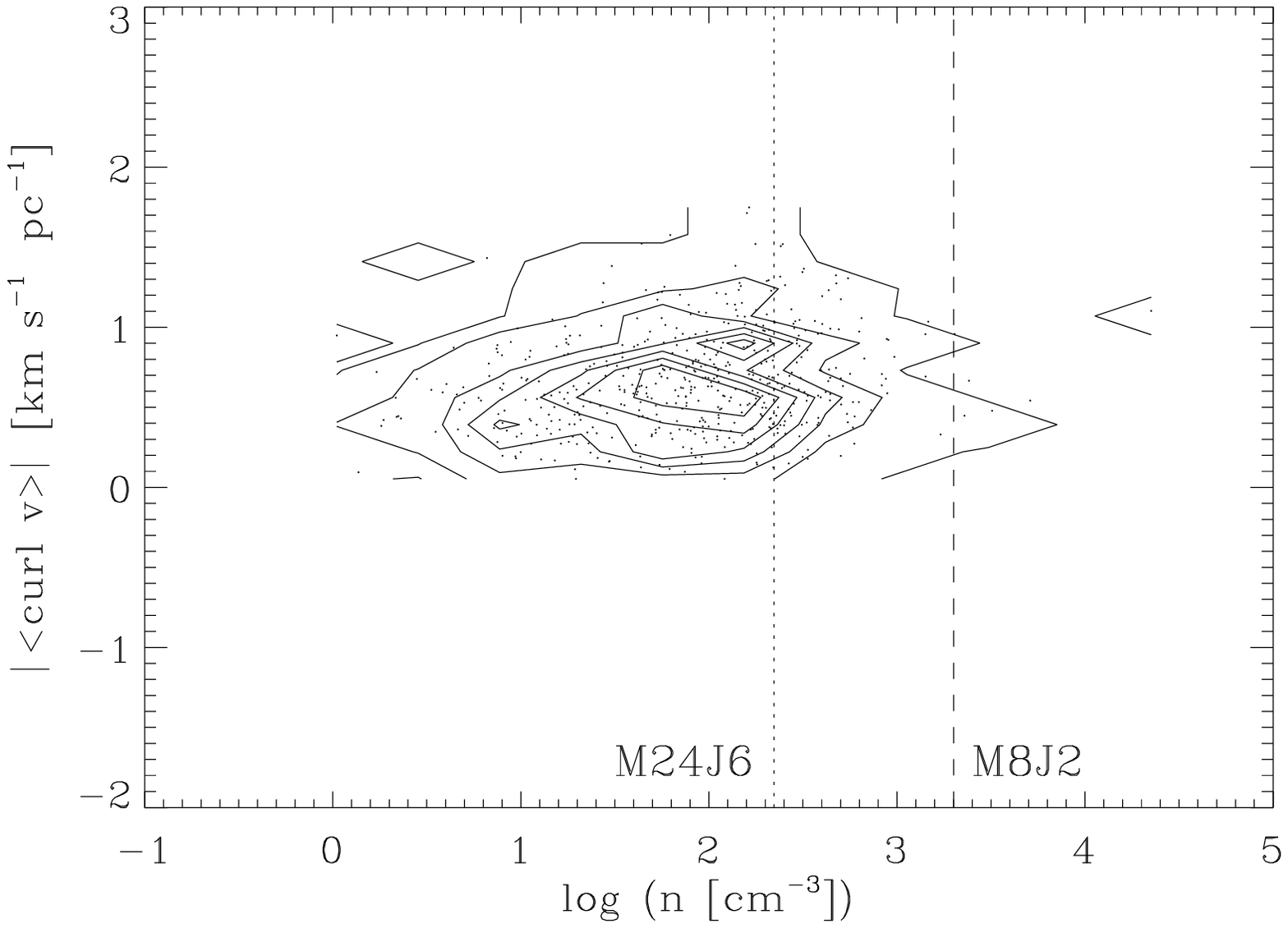} 
\caption{{  Same as Figs. \ref{fig:div_vs_ro_NSG} and
\ref{fig:div_vs_ro_SG}, but for the magnitude of the vorticity at $t=4$
Myr ({\it left panel}, without self-gravity) and $t=14$ Myr ({\it right
panel}, with self-gravity) (respectively, compare to the { {\it left
panels} of Figs. \ref{fig:div_vs_ro_NSG} and \ref{fig:div_vs_ro_SG}}). It
is seen that comparable magnitudes of 
the (negative) divergence and of the vorticity occur before and after
turning self-gravity on.}}
\label{fig:curl_vs_ro}
\end{figure*}

{ It is interesting to compare the magnitude of the convergence seen
in Figs. \ref{fig:div_vs_ro_NSG} and \ref{fig:div_vs_ro_SG} with that
of the vorticity, in order to get an idea of which type of motion
(potential or solenoidal) dominates within the subregions of run
Ms24J6. Again, we calculate the vorticity in Fourier space as 
\begin{equation}
{\cal F}(\nabla \times v) = -i {\bmath k} \times {\bmath v_k},
\label{eq:Four_curl}
\end{equation}
and use a similar averaging procedure for the Ms8J2-sized sub-boxes of
run Ms24J6 as for the divergence. Since the vorticity is a
(pseudo-)vector, in Fig.\ \ref{fig:curl_vs_ro} we plot the magnitude of
the average vorticity in each sub-box. This is shown at $t=4$ Myr (a
time without self-gravity; {\it 
left panel}) and at $t=14$ Myr (a time with self-gravity; {\it right
panel}). In general it is seen that the magnitudes of the 
divergence and of the vorticity are comparable, suggesting that the
sub-boxes contain comparable amounts of the two types of motion. 

Note, however, that it is not possible from these figures to obtain a
quantitative estimate of the amount of kinetic energy contained in each
type of motion, since, contrary to the case of the energies, { not} the
divergence nor the vorticity nor their sum are bounded or conserved in
any sense. It would be necessary to compare the kinetic energy in each
type of motion directly but, because this task is not straightforward,
we defer it { to} a future study. In any case, our simulations show that,
to order of magnitude, the potential and solenoidal parts of the
velocity gradient are comparable, and therefore we expect the fraction
of compressive energy present in dense regions to be substantial.}

\section{Star Formation Efficiency in Constant-Virial-Parameter
Structures} \label{sec:SFE}

In this section we now proceed to measure the SFE in terms of what KM05
called the star formation rate per free-fall time, \sfrff, in the three
simulations, in order to test whether the predictions of their model are
verified. { Following KM05, we define the \sfrff\ as the fraction of
a cloud's mass that is deposited into stars in one free-fall time.}  As
in previous papers \citep[e.g., ][]{VKB05, GVKB07}, we { approximate this
fraction as the fraction of the total mass in the simulation that is in
regions with density above a certain threshold. The threshold is chosen
to be unambiguously indicative that collapse has occurred.} 
Specifically, we take a threshold density $n_{\rm thr} = 1000~n_0$,
which is a much larger density than can be achieved through the
turbulent fluctuations alone in any of the runs, { and guarantees that
the mass contained in objects above those densities is either collapsing
or gravitationally locked within them and does not redisperse (except
for slight fluctuations, discussed in the next paragraph)\footnote{Note
that, in our simulations, gravitational collapse is necessarily
terminated when most of the collapsed mass is deposited into a single or
a few grid cells. It is in this sense that the material is ``locked''
into the collapsed objects, and by no means it should be interpreted as
implying that a real hydrostatic object (other than { one or several
stars}) has formed.}. As noted in
\citet{VKB05}, the collapsed (or ``accreted'') mass fraction does not
depend heavily on the chosen threshold, as long as the latter satisfies
the above conditions. As a confirmation, we have also used a threshold
$n_{\rm thr} = 500~n_0$, and found no significant difference in the
results, as described below. Note however that our procedure may
overestimate the actual mass deposited into protostellar objects, since
it assumes a 100\% efficiency of conversion of gas into stars above the
threshold density, while in actual cluster-forming cores, the expected
efficiency is closer to 30-50\%
\citep{LL03}.} 

\begin{figure*}
\includegraphics[width=0.45\hsize]{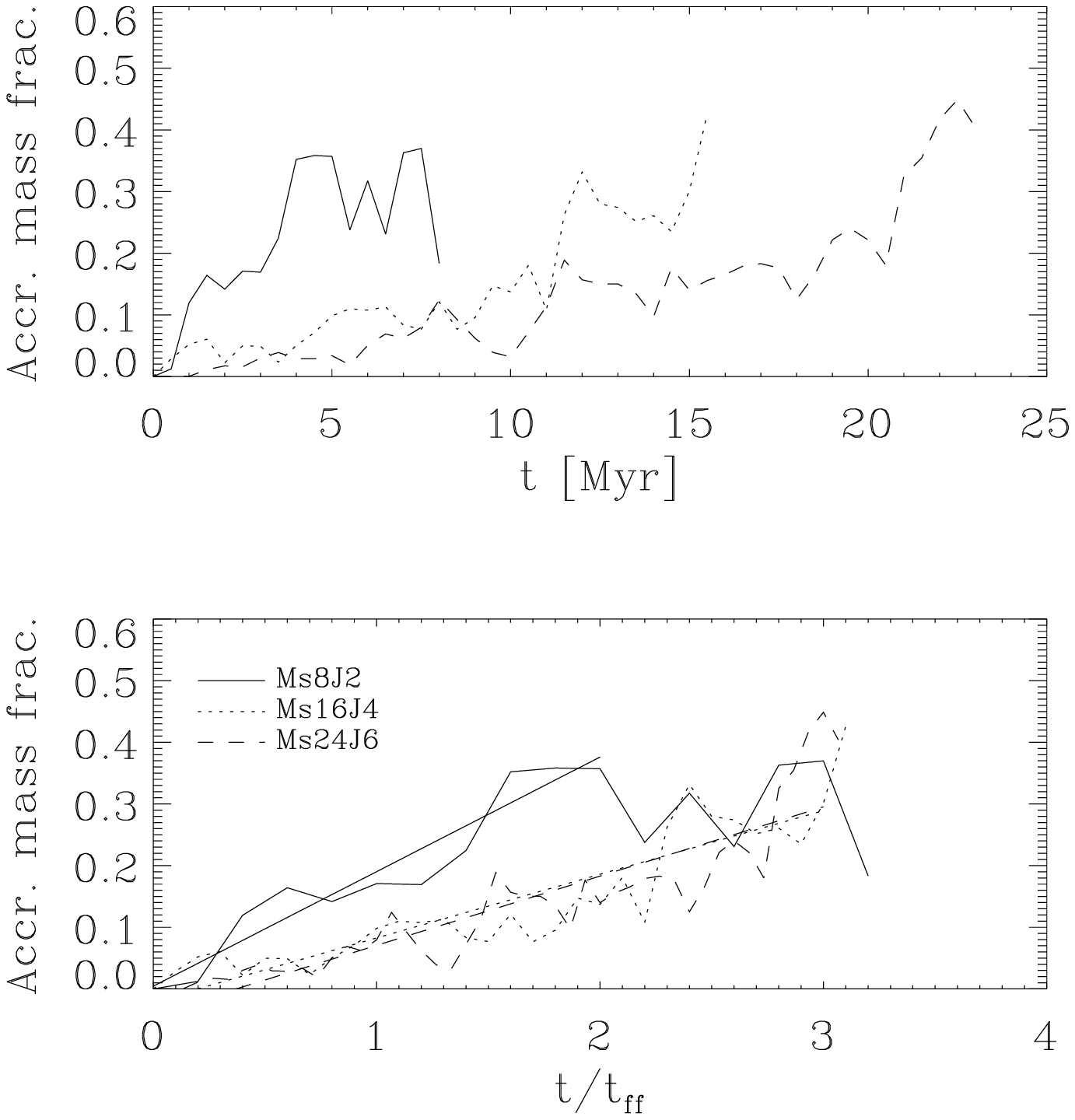}
\hspace{0.5cm}\includegraphics[width=0.45\hsize]{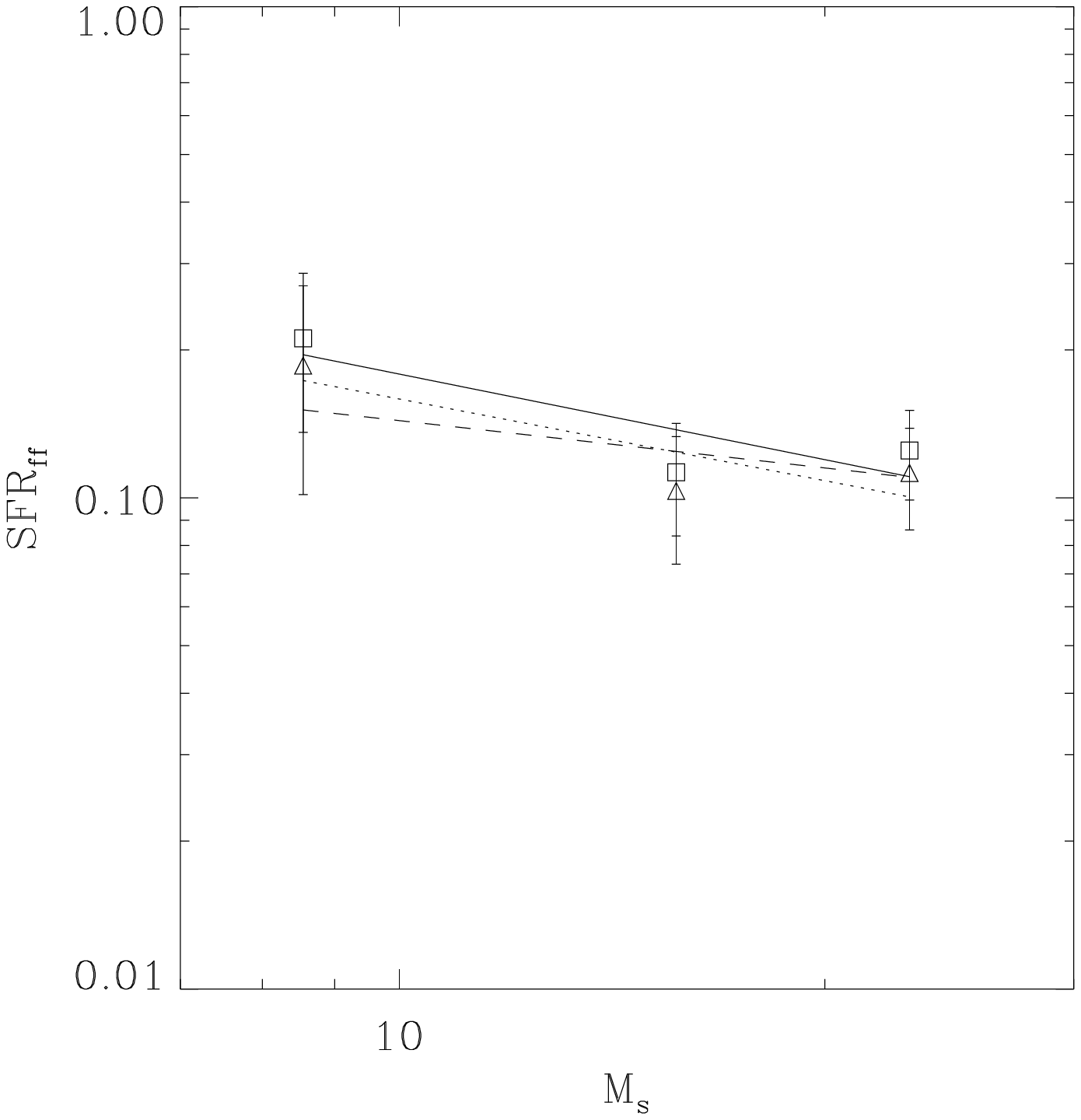}
\caption{{\it Left panels:} Fraction of mass accreted onto collapsed
objects as a function of time for the three simulations, with time in
units of Myr ({\it top left}) and of the free-fall time of each
simulation ({\it bottom left}). The average \sfrff\ is computed by
fitting a least-squares line to the plot of accreted mass vs.\
time-in-units-of-$\tff$. {\it Right panel:} Star formation rate per
free-fall time (\sfrff, i.e., average accreted mass after one free-fall
time) as a function of the time-averaged rms Mach number of the
simulation for the three runs. {\it Triangles} (resp. {\it squares})
denote results obtained with a density threshold $n_{\rm thr} =
1000~n_0$ (resp. $n_{\rm thr} = 500~n_0$) for defining the collapsed
objects. The {\it dotted} and {\it solid} lines show the best fits
through the data points for $n_{\rm thr} = 1000~n_0$ and $n_{\rm thr} =
500~n_0$, respectively. The slope of the {\it dotted line} is $-0.55$,
while that of the {\it solid line} is $ -0.58$.  The {\it dashed} line
shows the prediction of the KM05 model, with slope $-0.32$.}
\label{fig:macc}
\end{figure*}

 The left panels of Figure
\ref{fig:macc} show the evolution of the mass accreted onto collapsed
objects as a function of 
time after having turned gravity on, in units of Myr ({\it upper}
frame) as well as in units of the free-fall time $\tff$ ({\it lower}
frame). The \sfrff\ is then the slope of the collapsed mass fraction
versus time, the latter in units of $\tff$.

The accretion histories of the runs are seen to be noisy, { even 
exhibiting slight, occasional} drops. This reflects the fact that our
numerical scheme does not include any prescription for sink particles or
cells, and therefore the mass in the collapsed object is not fully
``locked'' in the object -- the outer parts of it may have densities that
oscillate { about} the threshold we use for defining the collapsed
objects. This implies that the \sfrff\ is determined only to within a
certain uncertainty in our simulations. For estimating the average
\sfrff\ { at each $\Ms$, we thus fit a least-squares line to the plot
of accreted mass vs.\ time (the latter in units of $\tff$; lower left
panel of Fig.\ \ref{fig:macc}), and show error bars corresponding to
$\pm 3 \sigma$, where $\sigma$ is the standard deviation of the
fit. Note also} that the accreted mass for run Ms8J2 appears to saturate
at $t \sim 5~\hbox{Myr} \sim 2 \tff$, so for this run we only take into
account times shorter than this saturation time.

With this procedure, we obtain the average values of the \sfrff\ shown
in the {\it right panel} of Figure \ref{fig:macc}. { For reference,
we show the results obtained with the two thresholds $n_{\rm thr} =
500~n_0$ and $n_{\rm thr} = 1000~n_0$. We see that the data points are
shifted slightly to higher values for the lower threshold, since the
structures contain more mass, but the trend of the \sfrff\ is
essentially the same in both cases.}

Our results can be compared with the 
prediction of the KM05 model, which those authors fitted by 
\begin{equation}
\hbox{\sfrff} \approx 0.014 \left(\frac{\alpha}{1.3} \right)^{-0.68} \left(
\frac{\Ms}{100} \right)^{-0.32}.
\label{eq:KM05_SFRff}
\end{equation}
Thus, a residual dependence on the Mach number $\sim \Ms^{-0.32}$ is
expected in the KM05 model even at constant $\alpha$. The slope of
$-0.32$ is indicated in the {\it right panel} of Figure \ref{fig:macc}
by the {\it dashed} line, while the {\it solid} and {\it dotted}
lines indicate fits to our data, with slopes $-0.58$ for $n_{\rm thr} =
500~n_0$ and $-0.55$ for $n_{\rm thr} = 1000~n_0$, respectively. 

We see
that, within the uncertainties, our results can accomodate the slope
predicted by KM05, although the agreement is only marginal, { because the
most probable slope that we measure is significantly steeper than
theirs. Besides, the
intercept of the fitted line appears to imply a larger coefficient
than the one in eq.\ (\ref{eq:KM05_SFRff}). For
example, at $\Ms = 10$, the {\it dotted} line ($n_{\rm thr} = 1000~n_0$)
in the right panel of Fig.\ \ref{fig:macc} gives \sfrff $\approx 0.15$,
while eq.\ (\ref{eq:KM05_SFRff}), with $\alpha = 1.2$, predicts \sfrff\ $=
0.031$. 

Several effects, both physical and of procedure, can account for these
differences. Possible effects of procedure are: a) We are assuming a
100\% efficiency of conversion from gas to stars above the density
threshold. Instead, this efficiency is known to be $\sim 30$\% for
high-mass star forming cores
\citep{LL03} and even lower for low-mass ones, so we are clearly
overestimating the total mass deposited into stars by a factor of
3-10.\footnote{{ Note that, because we use a threshold that is simply
a fixed 
multiple of the simulation's mean density, we may be biasing the measured
slope, since the {\it physical} threshold density is highest
in the small-scale run Ms8J2 and lowest in the large-scale one,
Ms24J6. However, correcting for this by, say, taking a fixed physical
density in all three simulations will actually tend to make the slope of
the \sfrff\ vs. $\Ms$ curve even steeper, since a higher threshold in
code units should be used in Ms24J6, leading to an even smaller measured
\sfrff. Thus, this effect cannot account for the fact that we measure a
steeper slope than that predicted by KM05.}} b) KM05 calibrated their
model using linear approximations to the SFEs of
\citet{VBK03}, while in reality the mass accretion histories presented
by those authors are strongly nonlinear. c) We have defined the
free-fall time simply as $\tff = \LJ/\cs$, which is larger by a factor of
$\sqrt{32/3 \pi}$ than the alternate definition $\tff^{\prime} \equiv (3 \pi/32
G\rho)^{1/2}$. Our choice is a reasonable
estimate, since actual collapse appears to take from $\sim 1.6$ to a few times
$\tff^{\prime}$ \citep{Larson69, GVSB07, GVKB07}, due to the non-negligible
role of the thermal pressure. However, KM05 used $\tff^{\prime}$
\citep{KT07}, so our measured 
masses are larger than those that would be measured out to $\tff$ by a
factor of a few, raising the intercept of our fit with respect to that
of KM05.

On the other hand, possible
physical effects are: a) the fact discussed in \S
\ref{sec:div_v} that the clumps in each simulation lack support in
comparison with the KM05 model because the motions are not fully random,
but instead involve a significant inflow component. b) Most importantly,
the mass that appears to be collapsing in our simulations is not that in
subsonic, super-Jeans regions, but rather the mass in regions in regions
that are supersonic but still gravitationally unstable. This allows for
a larger amount of mass involved in the collapse, since in general
regions with larger velocities are expected to be larger, and thus more
massive.  In conclusion, it is not possible to determine whether the
differences between the KM05 prediction for \sfrff\ and our results are
significant. However, it is clear that the mechanism of collapse acting
in our simulations is different from what they, as well as
\citet{Padoan95} and \citet{VBK03}, assumed, since there are no subsonic,
super-Jeans structures in our simulations.}

\section{Discussion and Comparison to Previous Work} \label{sec:discussion}

\subsection{Velocity convergence} \label{sec:disc_vel_conv}

Our results have a number of implications for our understanding of the
velocity field in MCs and for analytical models of star formation. A
first result is that in sub-boxes of the large-scale run (Ms24J6) of the
same size as the small-scale run (Ms8J2), the velocity field exhibits a
clear trend towards being convergent if the subregion is overdense with
respect to the rest of the simulation. This 
suggests that run Ms8J2, with its zero overall mean divergence, is not
representative of the typical region of the same size embedded within a
larger medium, since it lacks an expected mean divergence of $\sim - 0.6
\kms {\rm pc}^{-1}$, as predicted by eq. (\ref{eq:div_vs_ro}).
 
This result also suggests that a significant component of the observed
linewidths in MCs and their substructure should be compressive { (i.e.,
conform with the LSI scenario)}, in agreement with the original
suggestion of \citet{GK74}.  It is even more relevant that this result
is observed in numerical simulations of {\it driven} turbulence {
(with the driving applied at the large scales and being purely
rotational)}, contrary to the frequent belief that such large-scale
inflows occur only in decaying turbulence. It is necessary to stress,
however that our analyses cannot distinguish whether the convergent
motions in the overdense regions are a cause or a consequence of
gravitational contraction. This is beyond the diagnosing capabilities of
the studies we have performed.

Dimensionally, the values of the mean divergences we have found for the
sub-boxes of our 
large-scale simulation can be compared to the typical values of the
velocity gradients found by \citet{GBFM93} in dense cores of MCs, which
they however interpreted as representative of uniform rotation. For
cores with typical sizes between $\sim 0.1$ and 1 pc, those authors found
velocity gradients ranging between 0.3 and 4$\kms$ pc$^{-1}$. Our
average value of the velocity divergence for regions of { the
same size and density as those of run Ms8J2, $\nabla \cdot \bu \sim -
0.6 \kms$ pc$^{-1}$}, seems to be 
on the low side of this distribution, although it may seem reasonable if
we consider that only a {\it fraction} of the kinetic energy should
be compressive { (i.e., non-rotational)} in general. We can also compare
with Larson's (1981) 
linewidth-size relation, which, for $^{12}$CO and $^{13}$CO data reads
$\Delta v 
\approx 1 \kms [L/{\rm 1~pc}]^{1/2}$ \citep[e.g.,][]{Solomon_etal87,
HB04}. Thus, for $L\sim 1$ pc, Larson's relation predicts a velocity
dispersion roughly { 1.8 times} that of our ``typical'' velocity
convergence at the scale of run Ms8J2, given by
eq. (\ref{eq:div_vs_ro}), suggesting that, on 
average, the { velocity dispersion contains comparable amounts of
compressive and non-compressive energy}. Of course, the fluctuations are
large on both our velocity convergence-density relation
(eq. \ref{eq:div_vs_ro}) and on Larson's relation, so { large deviations
from this mean trend can be expected in individual clumps.}

Since the compressive part of the velocity works to promote compression,
the main implication of our result is that {\it not all of the
non-thermal kinetic energy in clouds and clumps is available for support
against gravity}, a fact that has been overlooked by analytic models of
star formation from the turbulent conditions in molecular clouds and
clumps. { In particular, the models by PN02 and KM05 assume that a
star-forming core must exceed the thermal Jeans mass in order to
collapse but, as shown by \citet{HF82}, this mass is reduced in the
presence of a compressive velocity field. The model by \citet{HC08} does
consider an additional ``support'' by the turbulence, but this may
again be reduced if the ``turbulence'' actually contains a significant
fraction of inwards motions.}

The LSI scenario is naturally motivated by the fact
that only compressive motions can 
produce density fluctuations, while vortical modes are
incompressible. Thus, if the density enhancements (clumps) in a flow are
produced by turbulent fluctuations, the velocity field within these
clumps { is likely to} still exhibit the signature of the external
convergent motions that 
formed them \citep{BVS99}. This scenario also has the implication that the
largest velocity dispersions are expected to occur not in the densest
parts of the structures, but at their outskirts, 
since the densest gas has been shocked and has slowed down
\citep{Klessen_etal05, GVSB07}. This effect has been found
observationally, albeit with { lower velocity dispersion than in the models}
\citep[see, e.g.,][and references therein]{Andre_etal08}. Finally, the
LSI scenario is also fully consistent with the
observation by \citet{HeBr07} that Principal Component Analysis of the
velocity field in molecular clouds and their substructure systematically
shows the dominance of a whole-cloud dipolar pattern.

At first sight, one could think of two possible
ways to avoid the LSI picture. One { possibility} would occur if the
clumps were quasi-static structures in single-phase media confined
by ram pressure \citep{BM92}. In this case, the boundaries must
be accretion shocks \citep{FW06, Whit_etal07, GVSB07}, so that the mass
of the structures must grow over time, possibly eventually becoming
strongly self-gravitating and proceeding to collapse. { Thus}, even though
such clumps can be quasi-static in their central parts, they must be surrounded
by an accreting envelope that involves a net convergence of the velocity
field \citep{GVSB07}. { The other possibility would occur } if the
medium, even within MCs, turns out to be thermally bistable
\citep{HI06}, in which case the clumps { could be hydrostatic objects
bound by the thermal pressure of their warm, tenuous environment, as in
the classic two-phase model of \citet{FGH69}.} However, even if {
molecular clouds are thermally bistable}, clumps that are at much higher
thermal pressures than their 
surroundings must be driven either by ram-pressure compressions or by
self-gravity { \citep{BVHK08}}. So, there appears to be no escape
from the need to have 
convergent flows involved in the formation and evolution of the densest,
star-forming clumps.

\subsection{Absence of subsonic, super-Jeans structures}
\label{sec:disc_no_ss_SJ}

Another result we have obtained is that, in our driven-turbulence
simulations, no simultaneously subsonic and super-Jeans structures were
found, either before or after self-gravity was turned on. Of course,
this result does not rule out the existence of such structures, and in
fact subsonic, super-Jeans cores are routinely observed \citep[e.g.,
][]{Myers83, Andre_etal07}. Our result may be an artifact of the
turbulence being continually driven and/or the absence of magnetic
fields and/or the fact that the actual values of $\alpha$ in our
simulations are somewhat greater than unity. Nevertheless, our
simulations produced abundant collapse, indicating that the formation of
simultaneously subsonic and super-Jeans structures \citep{Padoan95,
VBK03} is not the only possible route to collapse. Since this notion is
at the foundation of models such as those of PN02 and KM05, it is likely
that those models may need to be revised to consider the possibility
that stars may form via the collapse of larger-scale, supersonic
regions, in which the motions are not fully supportive against gravity.

It is also important to reconsider the results of \citet{VBK03} in the
light of our present results. In that paper it was shown that there
exists a correlation between the sonic scale of the turbulence and the
SFE, so that, as the sonic scale becomes smaller (at constant $J$), the
SFE decreases. This was interpreted as indirect evidence that the
available mass for collapse decreased as the sonic scale became smaller,
and therefore that the collapsed objects indeed originated from
subsonic, super-Jeans structures. However, our finding { that there exist
alternative routes to collapse that are} not based on the formation of
subsonic, super-Jeans structures suggests that the correlation found by
\citet{VBK03} may be simply indicative of a general scaling of both the
SFE and the sonic scale with rms Mach number, but not that the fraction
of mass in subsonic, super-Jeans structures directly measures the mass
that is on route to collapse at any given time.

\subsection{Role of gravity in the formation of Jeans-unstable
structures and the density PDF} \label{sec:grav_form_struct}

Our simulations also suggest that self-gravity is not only involved in
the {\it capture} of turbulent density fluctuations to make them
collapse, but also in the {\it production} of collapsing objects, as
shown by the distortion of the density 
PDF and by the increase of the fraction of self-gravitating regions at
small scales in the presence of self-gravity. In this sense, these
results may represent the driven-turbulence counterpart of the results
by \citet{CB05}, who concluded from decaying turbulence simulations that
the turbulence does not directly produce the collapse of clumps by making
them reach their own Jeans mass, but
rather just produces the seeds for subsequent gravitational
fragmentation of the large-scale gravitationally unstable structures.
In our driven case, it appears that turbulence alone produces only a few
super-Jeans structures, while, in the presence of
self-gravity, super-Jeans objects are much more readily produced. 

This result is likely to have an implication for the PN02
model. This model assumes that turbulence alone is responsible for the
formation of the cores, with gravity only playing a role if they become
Jeans unstable { and collapse}. Our results suggest
instead that turbulence alone is insufficient for producing a hierarchy
of structures whose mean densities scale with size in a
\citet{Larson81}-like way (recall our simulations are constructed to
have mean densities inversely proportional to their sizes, but
turbulence alone appears incapable of producing regions like run Ms8J2
within run Ms24J6). Turbulence must be considered in conjunction with 
self-gravity for the production of the denser structures.

{ This fact is also likely} to have an implication for the interpretation
of the recent findings by \citet{JM06}. These authors concluded from
non-self-gravitating simulations of the supernova-driven ISM with a
fixed imposed supernova rate, that this driving alone is insufficient to
sustain itself, as it does not deposit a sufficiently large amount of
mass in Jeans-unstable regions to be consistent with the imposed
supernova rate. Our result that gravity participates in the {\it formation}
of Jeans-unstable regions suggests that the discrepancy between the
applied supernova rate and the rate of production of Jeans-unstable
regions in their simulations they reported (roughly an order of
magnitude) may actually be an upper limit, so that supernova driving may
be { more efficient} in driving secondary star formation than they
concluded.

{ Finally, the} fact that the density PDF develops a high-density
tail in the 
presence of self-gravity also has implications. { Several models and
calculations \citep[e.g.,][]{Elm02, PN02, KM05}} rely on this
distribution, and it is necessary to 
assess the degree to which such a deviation may alter the results of
these models.

\subsection{Star formation rate per free-fall time}
\label{sec:disc_SFRff}

We have also measured the \sfrff\ in our three simulations,
all of which have turbulent driving applied at the largest scales in
each simulation, and approximately the same virial parameter $\alpha$,
{ thus matching the assumption made by KM05 to derive their equation
(30) (eq. [\ref{eq:KM05_SFRff}] in the present paper) that star-forming
clouds and clumps are all nearly virialized, with $\alpha \sim 1$ }. We found
the \sfrff s in our simulations to be marginally consistent
(i.e., within the range allowed by our uncertainties) with the
prediction of the model. This is not surprising, since our suite of
simulations was specifically designed to satisfy the {
$\alpha \sim 1$ assumption of KM05.} Note however
that, in spite of this specific design of the simulations, at face value
the slope of the \sfrff\ vs.\ $\Ms$ we find is larger than the
prediction by KM05. { This may be a reflection of our result} that
the sub-boxes of 
the large-scale simulation Ms24J6 with the same size and mean density as
the small-scale simulation Ms8J2 differed from it in that they contain a
non-zero mean convergence of the velocity field. This result contradicts
the {\it hypothesis} of the KM05 model \citep[and many others,
e.g.,][]{Elm02, MT03, MK04, HC08} that the turbulence at every scale is
completely 
random, and therefore acts as an isotropic pressure that provides
support against the self-gravity of the entire structure. If in reality
clouds and clumps contain a significant amout of kinetic energy in
convergent motions that do not oppose gravity, the dependence of the SFE
on $\Ms$ {\it at } $\alpha \sim 1$ might actually be somewhat steeper
than the prediction by KM05.  { This in fact may be the reason why our
simulations give a mean larger exponent in the relation between \sfrff\
and $\Ms$, since, even though each one of our simulations was
constructed as to have a random global turbulent velocity field, the
clumps formed self-consistently within them do not, on average.}

A final note of caution, however, is that our results have been obtained
in the simplest 
possible numerical setup, namely non-magnetic, isothermal media without
stellar feedback, and as such, cannot be considered definitive. We plan
to investigate thermally bistable, magnetized media with stellar
feeedback in future works.

\section*{Acknowledgments}
We thank Patrick Hennebelle for a critical reading of { an earlier
version of} the manuscript and
useful comments, and an anonymous referee for a thoughtful and thorough
report that much helped in improving the paper. The numerical
simulations were performed on the Cluster Platform 4000 (KanBalam) of
DGSCA, UNAM. This work has received financial support from grant
U47366-F (CONACYT) to E.V.-S., grants IN111606 and IN 117708 to R.F.G.,
IN110606 to 
J.B.-P., and 111006-3 to A.G. J.K. was supported in part by KOSEF
through the Astrophysical Research Center for the Structure and
Evolution of Cosmos and the grant of the basic research program
R01-2007-000-20196-0.

\bsp 

\label{lastpage} 

\end{document}